\documentstyle[aps,prl,tighten,eqsecnum]{revtex} 
 
\begin{document} 
\twocolumn
\narrowtext
\draft 
\date{\today}
\title{What is an atom laser?} 
\author{H.M. Wiseman} 
\address{Department of Physics, The University of Queensland, 
St.~Lucia 4072, Australia.}  
\maketitle
\begin{abstract}
An atom laser is a hypothetical device which would produce an 
atomic field analogous to the electromagnetic field of a 
photon laser. Here I argue that for this analogy to be meaningful it is 
necessary to have a precise 
definition of a laser which applies equally to photon or atom lasers. 
The definition I propose is based upon the principle that the output 
of a laser is well-approximated by a classical wave of fixed 
intensity and phase. This principle yields four quantitative 
conditions which the output of a device must satisfy in order for that 
device to be considered a laser. While explaining these requirements, I 
analyse the similarities and differences between atom and photon lasers. 
I show how these conditions are satisfied first by an idealized photon laser 
model, and then by a more generic model which can apply to atom lasers 
also. Lastly, I briefly discuss the current proposals for atom 
lasers and whether they could be true lasers.
\end{abstract}


\newcommand{\smallfrac}[2]{\mbox{$\frac{#1}{#2}$}}
\newcommand{\bra}[1]{\langle{#1}|} 
\newcommand{\ket}[1]{|{#1}\rangle}

\section{Introduction}

The field of quantum optics exists courtesy of the invention of the 
laser, now almost four decades ago \cite{SchTow58}. 
Not surprisingly, the mechanism of a laser 
is treated in most modern quantum optics textbooks 
\cite{Lou83,Hak85,MeySar90,Gar91,WalMil94} although 
there are exceptions \cite{VogWel94}. However, in none of these books 
is to be found an actual definition of the term ``laser''.
The expansion of the acronym itself --- light amplification by the 
stimulated emission of radiation --- falls well short of explaining 
what is special about laser light. For example, stimulated 
emission may dominate spontaneous emission even in a laser below 
threshold, and yet 
at least the quantum optics community believes that operation above 
threshold is essential for a device to be considered a true laser 
\cite{Lou83,WalMil94}.
The precise definition of a laser would clearly be a difficult task,
 so the approach by the authors of these books --- to present
 an analysis of the dynamics of some more or less 
idealized model of a laser, and then to look at the consequences of 
these dynamics on the properties of the light --- is quite 
understandable.

As well as being difficult, constructing a definition of a laser 
might be thought 
futile. One reason is that different groups of people 
(whether in theory, experiment, or applications) require quite different 
properties from a laser, and hence would differ in 
their own intuitive understanding of what makes a laser special. 
Another reason is that the same physics often 
 underpins the operation of lasers in vastly 
different regimes, so it would seem unnatural to exclude some 
regimes by saying ``this is no longer a true laser''.

These are powerful arguments against attempting a definition of the 
term ``laser''. However, a new field of physics has recently arisen which 
seems to need such a definition if its future direction is to be clear. 
This is the field of atom optics and, more particularly, atomic Bose 
condensation. Atom optics is the study of atoms under conditions where 
their de Broglie wave nature becomes important, suggesting an 
analogy with photons \cite{AdaSigMly94}. The recent experimental 
achievement of dilute gases of atoms which are highly Bose-degenerate
\cite{And95,Bra95,Dav95} has now deepened the analogy between photons 
and Bose atoms by demonstrating that their quantum statistics are 
identical. It is generally recognized 
 that the next logical step would be to try  
to create an atom laser, ``the equivalent of a laser for atoms and 
\ldots a source of coherent matter waves'' \cite{BurJulSuo96}. 
But in what sense is an 
atom laser to be equivalent to a photon laser, and what are the 
properties which make matter waves coherent?

Recently there have been a number of proposals for building a device
under various names similar to that of `atom laser'
\cite{WisCol95,SprPfaJanWil95,OlsCasDal95,GuzMooMey96,%
WisMarWal96,Hol96,MoyHopSav96}. Very recently
Mewes {\em et al.} \cite{Mew96}  have actually built an output coupler for a 
Bose condensate which they say ``can be regarded as a pulsed `atom laser'''. 
Although their experimental results are an important milestone, 
their cautious language makes it clear that the output of their device
is still a long way short of the coherent matter waves 
one would envisage for a true atom laser.
When a true atom laser is built, perhaps along the lines of one of 
the theoretical proposals mentioned above, it may be that it has as many and 
varied applications as a photon laser. In that case, the users of 
atom lasers will define what they mean by that term. But at present
 the applications for an atom laser are still
speculative. For this reason it seems to me that if 
experimentalists wish to build a true atom laser, there must be a fixed 
goal at which they can aim. That is to say, what is required is a 
definition of  
exactly what would constitutes an atom laser. In order to construct 
such a definition, it is first necessary to answer the question of
what constitutes a photon laser. 
If the term `laser' is to have any sensible meaning, 
its defining features should not depend on the `substrate' of the 
laser (atoms or photons). In particular, the acronym referring to the 
amplification of radiation is of no use.

In this paper I present a definition of a laser which applies equally 
as well to photon lasers as to atom lasers. In its most concise form, the 
definition is that the output of a laser should be well approximated 
by a classical wave of fixed amplitude and phase. This definition 
is clearly based on fundamental considerations rather than practical 
applications. A consequence of this is that some devices which are 
usually called lasers (such as pulsed lasers) would not be so termed under my 
definition. Of course I do not think that 
the community at large should change its vocabulary.
However, if the creation of an atom laser is to represent a 
really significant advance beyond the creation of a Bose condensate of 
atoms, then I think that the workers attempting to do this must accept some 
definition of a laser similar to that presented here. 

In the atom laser models which have been published so far 
most authors 
\cite{SprPfaJanWil95,OlsCasDal95,GuzMooMey96,MoyHopSav96} 
have not attempted to define what they mean by a laser. 
Rather, they  
have followed the course of the text-books mentioned 
above, in presenting the model for their atom laser, and analysing its 
dynamics. However, they generally fail to go on to analyse the 
properties of the output field.  
This seems to me to miss the most important point, because ultimately 
it is the output field of a laser which is actually used. 
The internal state of a laser is not particularly  
interesting in isolation: it comes quickly to a steady-state which is a roughly 
Poissonian mixture of number states (be it number of photons or atoms). 
Such a state has already been 
created in the atomic Bose condensates, so there is no need to mimic
the complex dynamics of a laser in order to reproduce this state. 
Moreover, some of these 
analyses \cite{SprPfaJanWil95,OlsCasDal95,MoyHopSav96} begin with a 
rate equation approach. While this is sensible for obtaining 
preliminary results, it precludes any possibility of calculating 
the first-order coherence of the laser output.

In Ref.~\cite{Hol96}, Holland {\em et al.} do claim to define a las\-er in terms 
of the ``essential properties of [its] output field''. 
However, the properties listed refer not only to the output field but also to 
the internal state of the laser, as they include the statement that 
the linewidth should be 
inversely proportional to the intracavity boson number.
It seems to me that the exact meaning of the statement is unclear: 
what is held fixed when the linewidth changes inversely with the 
change in the intracavity photon number? If the bare cavity linewidth 
is held fixed, then it is true that the laser linewidth for an ideal 
laser will vary inversely with the mean intracavity boson number, as 
will be shown in Sec.~IV. Unfortunately, 
the atom laser model presented in Ref.~\cite{Hol96} has other 
contributions to its linewidth apart from this fundamental one 
\cite{WisMarWal96}. Thus their model 
explicitly fails to satisfy their criterion, as would most optical lasers.

By contrast, 
the definition to be presented here is unambiguous and is
based purely on the properties of the laser output beam;  
the laser itself is treated as a black box with one relevant output. 
The only condition on the inputs is that they cannot be laser beams 
themselves, at least not of the same substrate (photons or atoms) as 
that beam which the device itself is supposed to produce. This 
concentration on the laser output does not imply 
that there are not analogies between the dynamics of typical photon lasers and 
those of suggested atom lasers, 
 nor that these analogies are not 
worth examining (in fact I will examine them in this paper). 
However, as will become apparent, the nature of the output field of 
the laser is sufficient to define what we mean by a laser. 
Some parts of this 
definition have been published before \cite{WisCol95,WisMarWal96}, but 
it has not been presented as a coherent whole.

This paper is organized as follows. The larger part is Secs.~II and 
III, in 
which I propose a set of four criteria for deciding whether a device can be 
considered a true laser.  Sec.~II contains the elementary 
definitions (which can be understood from the single-particle 
viewpoint) and Sec.~III contains the more interesting definitions 
which require quantum statistics to formulate. 
At the same time as giving this definition I 
will discuss their consequences for atom and photon lasers.
 In Sec.~IV, I present an extremely idealized  model of a 
photon laser to show how the required properties can be satisfied in 
principle. In Sec.~V, I present a more realistic model, which 
could apply to either a photon laser or an atom laser. This 
enables an identification of the typical dynamics which enable a 
laser to produce its 
characteristic output beam. I discuss how this model applies to 
photon lasers and the proposed atom lasers. Sec.~VI summarizes the 
most important points.

\section{Elementary Definitions}

\subsection{Directionality}

As stated in the introduction, the premise of this paper is that a 
laser should be defined solely in terms of its output beam. By 
stating the premise in this way, I have already made an assumption, 
namely that the output of a laser is a beam. That is to say, the 
first condition on a laser is that
\begin{description}
\item[(1)]{The output is highly directional.} 
\end{description}
This assumption allows us to separate out a longitudinal direction 
(the direction of propagation) and two transverse directions 
(the directions of diffraction). It is {\em not} implicit in this condition 
that the output of the laser must propagate in free space. Even for 
optical lasers, it is common now for the output to propagate in an 
optical waveguide such as a fibre. For an atom laser, a waveguide such 
as proposed in Refs.~\cite{SavMarZol93,MarSavZolRol94} may 
be essential. This is because, unlike photons, atoms do not naturally 
travel at the speed of light. In fact, the `natural' speed of atoms 
at the temperatures of the achieved Bose condensates is a few cm/s. At these 
speeds, the effect of gravity on the atomic motion is far from 
negligible over a macroscopic distance. As well as supporting the 
atoms against gravity, a waveguide could prevent spreading of the 
atomic beam due to diffraction. In addition, it would limit the number 
of transverse modes in the beam. In the ideal case, there would only 
be one transverse mode. While this is not an essential requirement for 
a laser, it is a desirable one, so I will list it as
\begin{description}
\item[(1$^+$)]{Ideally, the output has a single transverse mode.}
\end{description}
A single transverse mode also implies a single polarization state 
for photons and a single electronic state for atoms.

\subsection{Monochromaticity}

A second property of a laser output which can be stated in 
single-particle terms is 
that of monochromaticity:
\begin{description}
	\item[(2)]   The longitudinal spatial frequency of the output beam has a 
small spread in the sense that $\delta k \ll \bar{k}$.
\end{description}
The uncertainty in the spatial frequency is the reciprocal of
the characteristic coherence length $\ell_{\rm coh}=1/\delta k$. 
Thus the condition \textbf{(2)} can be expressed as
\begin{equation}
\ell_{\rm coh} \gg \lambda,
\end{equation}
where $\lambda = 2\pi/\bar{k}$ is the mean wavelength. Because atoms 
travel nonrelativistically, for this condition to be meaningful it is 
necessary to state the reference frame in which the spatial 
frequency $k$ is defined. This is because in a Galileian transformation 
to a frame moving 
at fixed speed $u$ the spatial frequency of a de Broglie wave 
is transformed to $k' = k - 
Mu/\hbar$, where $M$ is the mass of the atom. The obvious reference 
frame to choose is that in which the laser itself is at rest. 

For photons, there is no difficulty in reformulating condition {\bf 
(2)} in terms of the frequency spread or spectral width $\delta \omega = 
c\delta k$ being small 
compared to the central frequency $\bar\omega = c\bar{k}$. For atoms we have to
be careful about what frequency we are referring to. This is because 
unlike a photon field (the electromagnetic field), the atom field is 
unobservable, as will be discussed later. 
In particular, its global phase has no physical 
meaning, so its frequency is arbitrary. For instance, we could choose 
to include the rest mass $M$ in the energy which would give $\hbar\omega = 
Mc^{2}+ p^{2}/(2M)$, or we could choose to leave it out. 
The most physically 
sensible definition is to use only the kinetic energy, so that 
\begin{equation}
\bar\omega = \frac{\hbar 
\bar{k}^{2}}{2M},\;\;\frac{\delta\omega}{\bar\omega}=2\frac{\delta 
k}{\bar{k}}.
\end{equation}
This allows the monochromaticity condition to be stated as $\delta \omega \ll 
\bar\omega$ for the atomic case also.

\subsubsection{Dispersion}

As stated above, an important difference between photons and atoms is 
that the latter do not travel at a fixed speed. That is to say, a beam 
of atoms will disperse even in a vacuum \cite{ign}. It is important 
not to confuse dispersion with diffraction. The former is a longitudinal 
spreading of an atomic wavepacket, the latter a transverse spreading 
which is present also for photons. Although dispersion of an atomic beam 
will occur, it will be negligible over distances much shorter than a 
characteristic length which I will call the {\em dispersion length}. 
The dispersion length can be defined to be the propagation length over 
which the spread in an atomic wavepacket due to dispersion becomes
comparable to the original spread in the wavepacket due to the 
uncertainty principle. One would expect that for a laser this dispersion length 
would be much larger than the wavelength of the output.
For photons this is trivially satisfied, so below we need consider only 
atoms (or massive particles in general).

Let us assume that the momentum distribution in the output is Gaussian 
with width $\hbar\delta k$. Then we can imagine that each individual 
atom has a wavepacket which begins as a real Gaussian wavepacket in 
$x$ with $\delta x = 1/(2\delta k)$.  Then 
for Gaussian wavepackets it is easy to show 
that the {\em dispersion time} (the time for $x$-variance of the 
wavepacket to double) is given by
\begin{equation}
\tau_{\rm disp} = \frac{2M(\delta x)^{2}}{\hbar} = 
\frac{M}{2\hbar(\delta k)^{2}} = 
\frac{\bar\omega}{(\delta \omega)^{2}}.
\end{equation}
The dispersion length is found by multiplying by the mean velocity
\begin{equation}
\ell_{\rm disp} = \tau_{\rm disp}\frac{\hbar \bar{k}}{M} = 
\frac{\bar{k}}{2(\delta k)^{2}}.
\end{equation}
Since the wavelength of the output is given by $2\pi/\bar{k}$, the 
dispersion length will be much greater than a wavelength as long as 
$ \delta k \ll \bar{k}$. This is already guaranteed by the 
monochromaticity condition {\bf (2)} above. Also, the dispersion length will be 
much greater than the coherence length. This can be summarized as 
follows:
\begin{equation}
\ell_{\rm disp} \sim \frac{\bar{k}}{(\delta 
k)^2} \gg \ell_{\rm coh} \sim \frac{1}{\delta k} \gg \lambda \sim 
\frac{1}{\bar{k}}.
\end{equation}

\subsubsection{Longitudinal acceleration}

In the absence of relativistic effects, the frequency of a beam of 
photons is fixed. The same is not true of a beam of atoms. For 
example, a wave guide for an atomic beam could be directed so as to 
gradually lower the atoms a distance $d$. This causes the atoms to gain kinetic 
energy. Thus we have $\bar\omega' = \bar\omega + Mgd/\hbar$, but 
$\delta\omega' = \delta \omega$ so the emerging beam is more 
monochromatic than it was at the laser output. If we define $\zeta^{2} = 
\bar\omega' / \bar\omega$, then the effect of this longitudinal 
acceleration on the spatial frequency is
\begin{equation}
\bar{k}' = \zeta \bar{k}\;,\;\; \delta k' = \delta k / \zeta
\end{equation}
From this we find that the coherence length has increased both relative to the 
(decreased) wavelength and absolutely 
\begin{equation}
\ell_{\rm coh}'/\lambda' = \zeta^{2} \ell_{\rm coh}/\lambda\;,\;\;
\ell_{\rm coh}' = \zeta \ell_{\rm coh}
\end{equation}
and the dispersion length has increased both absolutely and relative 
to the (increased) coherence length:
\begin{equation}
\ell_{\rm disp}' = \zeta^{3}\ell_{\rm disp}\;,\;\;
\ell_{\rm disp}'/\ell_{\rm coh}' = \zeta^{2}\ell_{\rm disp}/\ell_{\rm 
coh}.
\end{equation}

Despite the plasticity of these features of an atom laser, it is 
still necessary for the monochromaticity condition {\bf (2)} to hold at the 
output of the laser. This is because we have assumed that the 
dispersion is still negligible after the longitudinal acceleration, 
which would be possible only if the original beam were approximately 
monochromatic.

\section{Quantum Statistical Definitions}

\subsection{Quantum field theory}

In this section we introduce definitions which require a 
many-body description of the output beam. The best description is that 
using quantum field theory, which I will now briefly outline

\subsubsection{Photon field}

Photons are the quanta of the (first) quantized electromagnetic 
field (see for example Ref.~\cite{Har72}). 
Under conditions {\bf (1)}, {\bf (1$^+$)}, and {\bf (2)} 
we need 
consider only longitudinal modes of the electromagnetic field 
propagating away from the laser with spatial frequency
 $k$ in the region of $\bar{k}$. If we denote the annihilation 
 operator for each longitudinal mode $a(k)$, then the canonical 
 commutation relations are
\begin{equation}
[{a}(k),{a}^\dagger(k')] = \delta(k-k'),
\end{equation}
and the Hamiltonian for these field 
modes is
\begin{equation}
H = \int_{0}^{\infty} \!\!d{k}\, ck {a}^\dagger(k) {a}(k),
\end{equation}
where $c$ is the speed of light. This is simply interpreted if $a^\dagger(k) 
a(k)$ is recognized as the operator for the density of photons in $k$ 
space. The evolution of
these operators in the Heisenberg picture is simply $a_{t}(k) = 
e^{-ickt}a_{0}(k)$. 

The field modes labelled by $k$ are completely delocalized. For the definitions 
which follow it is more convenient to have localized field 
operators, which we can define as
\begin{equation} \label{loc}
\tilde{a}(z) = 
\int_{\bar{k}-\Delta k}^{\bar{k}+\Delta k}\!\!\!\!d{k}\, a(k)e^{ikz},
\end{equation}
where $\delta k \ll \Delta k \ll \bar{k}$.
These operators obeys the approximate commutation relations
\begin{equation}
[\tilde{a}(z),\tilde{a}^\dagger(z')] \approx \delta(z-z').
\end{equation}
The actual width of this $\delta$ function is of order $(2\Delta 
k)^{-1}$ because $2\Delta k$ is the range of the integration. This 
spatial resolution is at least of order 
a wavelength $2\pi/ \bar{k}$, because $\Delta k \ll \bar{k}$. 
This commutation relation indicates that the operator $a^\dagger(z) a(z)$ 
can be thought of as the density of photons in real (one-dimensional) space,
provided one does not try to probe that space on a scale of order a 
wavelength or finer.
 
The evolution of these operators in the Heisenberg picture is simply
\begin{equation} \label{simHei}
\tilde{a}_{t}(z) = \tilde{a}_{0}(z-ct).
\end{equation}
Thus we could just as 
well parameterize this local field operator in time rather than space 
by defining
\begin{equation}
b(t) \equiv \sqrt{c}\tilde{a}_{t}(z_{1}) = \sqrt{c}\tilde{a}_{0}(z_{1}-ct),
\end{equation}
where $z_{1}$ is some convenient point in the output.
The prefactor $\sqrt{c}$ 
is so that these operators obey the commutation relation
\begin{equation} \label{bcr}
[b(t),b^\dagger(t)] = \delta(t-t'),
\end{equation}
where again the width of this $\delta$ function must be at least of 
order an optical period. With this proviso in mind, the operator 
\begin{equation}
I(t) = b^\dagger(t)b(t)
\end{equation}
 represents the density of photons in time, or the {\em photon flux} 
 at the position $z_{1}$.

\subsubsection{Atom field}

Atoms are composed of elementary fermions: quarks and electrons. An 
atom which comprises an even number of elementary fermions (which 
is to say an atom containing an even number of neutrons) will, under 
dilute conditions, behave like a particle which is a boson. That is to 
say, a dilute gas of such atoms has the same statistics as a gas of 
photons. However, there are important differences between these two 
gases. Photons are gauge bosons which carry the electromagnetic 
force, and arise from quantization of the electromagnetic field. There 
is no analogous force for which atoms are the carrier, and hence there 
is no fundamental atom field which when quantized yields atoms. 
Rather, atoms are particles, and the fundamental 
description for a system with many indistinguishable atoms is the appropriately 
symmetrized many-particle wavefunction. However, this formalism is 
very unwieldy and  inefficient. For this reason we introduce a quantum 
field description of the many atoms, by a process commonly called second 
quantization \cite{Har72}. Thus the atom field is secondary to the 
atoms, whereas the electromagnetic field is primary to the photons.

The end result of the second-quantization procedure
 is that the single particle Schr\"odinger 
wavefunction is transformed into a set of operators $\psi(z)$ 
parameterized by position, such that 
$\psi^\dagger(z)\psi(z)$ is the operator for the (linear) 
density of particles at position $z$. In this case, the commutation 
relations for Bose particles 
\begin{equation}
[\psi(z),\psi^\dagger(z')] = \delta(z-z')
\end{equation}
is exact. However for atoms it must be remembered that they can only 
be treated as bosons if the interatomic separation is large compared 
to an atomic radius, but this has already been assumed \cite{ign}. The 
field $\psi(z)$, although analogous to the 
electromagnetic field for photons, is different in that it does not 
carry any force, as explained above. 
This is a consequence of the conservation laws for elementary 
fermions, which imply that no Hamiltonian can be linear in a matter 
field such as $\psi$. Thus the field $\psi$ is strictly unobservable, 
although a bilinear combination such as the density $\psi^\dagger \psi$ is 
certainly observable.

Now consider an atom laser, and in particular a region of its output around some 
point $z_{1}$ where there is no longer any longitudinal acceleration, 
and where dispersion is still negligible. The field operator $\psi(z)$ 
may be decomposed  using the single-particle energy 
eigenfunctions $u_{k}(z)$ as
\begin{equation}
\psi(z) = \int_{-\infty}^{\infty}\!\!d{k}\, a(k) u_{k}(z),
\end{equation}
where $k$ is the spatial frequency at the point $z_{1}$. 
The modes $u_{k}(z)$ will not necessarily be of the form $e^{ikz}$ because we 
have allowed for the possibility of longitudinal acceleration prior to 
the point $z_{1}$. Nevertheless, a decomposition is possible in 
principle. It is therefore also possible to recompose these 
delocalized operators into a new $\tilde{a}(z)$ which contains only 
those $a(k)$ in the region of $\bar{k}$ (the mean spatial 
frequency of the output at the point $z_{1}$), as was done for the photon 
case (\ref{loc}). 

Because of the possibility of longitudinal acceleration, we cannot 
write down as simple an expression as (\ref{simHei}). Nevertheless the 
assumed negligibility of dispersion up to $z_{1}$ means that we can 
write 
\begin{equation}
\tilde{a}_{t}(z_{1}) = \tilde{a}_{t_{0}}(z_{1}-\xi(t-t_{1})),
\end{equation}
for some function $\xi(t)$ which satisfies $\xi(0)=0$ and $\dot{\xi}(t) > 0$. 
This again means that we can 
parameterize the output using the time variable by
\begin{equation}
b(t) = \sqrt{c}\tilde{a}_{t}(z_{1}).
\end{equation}
Here $c = \hbar \bar{k}/M = \dot\xi(0)$ is the mean speed of the atoms at $z_{1}$. 
We can then calculate the commutation relations 
\begin{eqnarray}
[b(t),b^\dagger(t')] &=& c[\tilde{a}_{t}(z_{1}),
\tilde{a}_{t}^\dagger(z_{1}-\xi(t'-t))]  \nonumber\\
&=& c\delta(\xi(t'-t)) \,=\, c\delta(t-t')/\dot{\xi}(0) \nonumber\\
&=& \delta(t-t')
\end{eqnarray}
to be the same as for the photon case (\ref{bcr}), providing the 
width of this $\delta$ function is again taken to be large compared 
to $\hbar$ divided by the mean kinetic energy. 
Thus $I(t)=b^\dagger(t)b(t)$ can be interpreted as  
the approximate atom-flux operator at $z_{1}$.

\subsection{The fundamental principle}

Having established that the output field of a laser can be represented 
by a field operator $b(t)$ satisfying approximate $\delta$-function commutation 
relations, I can now state the fundamental principle behind the 
criteria for deciding whether a device can be considered a laser. 
That principle is that
\begin{description}
	\item[($\pi$)]  The output is well-approximated by a classical wave of 
	fixed intensity and phase.
\end{description}
In slightly more mathematical terms, we require
\begin{equation} \label{pi}
b(t) \approx \beta(t) \equiv \beta e^{-i\bar{\omega}t},
\end{equation}
where $\beta$ is a $c$-number and the meaning of the $\approx$ sign will 
become clear in the remainder of Sec.~III. The fact that $\beta(t)$ has only 
trivial time-dependence indicates that I am here assuming that the 
output of the laser must be a {\em stationary process}. 
This excludes pulsed lasers from the class of true lasers. This is 
necessary if there is to be a distinction between projecting an 
atomic Bose-condensed gas across the laboratory and constructing a 
true atom laser.

The zeroth order approximation $b(t)=\beta(t)$ is used  
in quantum optics whenever the driving of a system (such as an atom or 
a cavity) by a laser is treated as classical driving. Assumption 
{\bf ($\pi$)} is also 
necessary for the first order (linearized) approximation of the 
intensity fluctuations of a laser by
\begin{equation}
\beta^{*}(t)\delta b(t) +  
\beta(t)\delta b^\dagger(t),
 \end{equation}
where $\delta b = b- \beta$. Thus the linearization of the fluctuation 
in an atomic travelling-wave field is only permissible if that field 
is the output of an atom laser.

\subsubsection{Mean amplitude?}
 
The first obvious requirement for $b(t)$ 
to be approximated by $\beta(t)$ would seem to be
\begin{equation}
\langle b (t) \rangle = \beta(t).
\end{equation}
Unfortunately, this
cannot be true for an atom laser. As explained above, atoms are particles of 
matter, rather than gauge bosons like photons. Thus, the atom field is 
not an observable like the photon field. Only bilinear combinations of 
the atom field are observable. All single-atom quantities, such as the 
atomic dipole, or the atomic momentum, correspond to field observables 
which are bilinear in the atom field. Similarly, no 
Hamiltonian is linear in the atom field, because this would imply the 
creation of atoms out of nothing. Of course, it is possible to create 
atoms in a particular electronic state from atoms in a different 
electronic state. At higher energies, it is possible to create atoms by 
splitting up molecules. At still higher energies, it is possible to create atoms 
out of pure energy by simultaneously creating an anti-atom. But all 
of these process involve other matter fields, and the fundamental Hamiltonian 
will always be bilinear in the total matter field which includes all 
elementary particles in the universe \cite{ManSha93}. Thus no interaction can 
produce  a quantum state in which the mean value of a matter field is 
different from its initial value of zero. We can therefore conclude that 
\begin{equation} \label{meanzero}
\langle b(t) \rangle = 0
\end{equation}
for an atom laser. 

In contrast to an atom field, the electric field 
{\em is} an observable, and it is linear in $b(t)$, so in principle 
an optical laser can have a mean field. However,
in an optical laser starting from the vacuum, 
the phase of the field produced 
cannot be predicted unless it is seeded with another laser. Thus in a 
strict sense the expected value of the mean field of an optical laser is 
zero also, because the average of $e^{i\phi}$ over all phases $\phi$ 
is zero. One could argue that the actual value of $\phi$ is found by 
making a phase-sensitive measurement of the output of the laser. 
The problem is that at optical frequencies, such measurements are almost always 
made relative to the phase of another laser beam, so they do nothing 
to establish the absolute phase of the electromagnetic field. 
Nevertheless it is possible in principle to determine this phase 
absolutely, for example by shooting an electron through a specific 
point in an electromagnetic standing wave at a particular time, and 
seeing in which direction it is deflected due to the Lorentz force. 
This would require timing to better than $10^{-16}$ seconds, which is 
why in practice only relative phase is measured at optical frequencies.
For a more technical discussion of the
differences and similarities between photon and matter fields, 
see the Appendix.

In summary, the zero mean field result (\ref{meanzero}) is true for 
both photon and atom lasers. For a photon laser we can 
believe that the output really is in a state with a well-defined 
amplitude and phase, but we don't know what that phase is. 
In this view, the result (\ref{meanzero}) is to be understood as a
 classical average over all possible phases. 
For an atom laser this is the wrong physical picture because a state 
with a mean field amplitude is physically impossible and the absolute 
phase of an atom field is absolutely unobservable. Nevertheless, 
this picture does not lead to contradictions, and may even be useful. 
Indeed, the imagining 
of the existence of a well-defined phase for a matter field is the 
essential element in the spontaneous symmetry breaking hypothesis in 
superfluidity and superconductivity
which has served the condensed matter physics community 
well for many decades \cite{And84}. However, it is not necessary to 
use this non-physical hypothesis in order to define a laser, and 
therefore I will not use it. 

\subsection{Well-defined intensity}

Even if the mean amplitude of a laser is zero, its mean intensity 
certainly is not. Thus the first requirement for approximating $b(t)$ 
by $\beta(t)$ is that
\begin{equation}
\langle I(t)\rangle= \langle b^\dagger(t) b(t) \rangle = |\beta|^{2}.
\end{equation}
However, to justify the $\approx$ sign in Eq.~(\ref{pi}) we require 
more than an equation of mean values. We also require that the 
fluctuations in the intensity be small in some sense. One's first 
thought might be to require that the standard deviation in $I(t)$ be 
small compared to its mean. Unfortunately, $\langle I(t)^{2}\rangle 
\to +\infty$, because of the $\delta$-function commutation relations 
(\ref{bcr}). For this reason it is necessary instead to consider the 
two-time correlation function for $I(t)$, and split off the singular 
part:
\begin{eqnarray}
\langle I(t+\tau) I(t) \rangle &=& \langle b^\dagger(t+\tau)b(t+\tau) b^\dagger(t) 
b(t) \rangle \nonumber \\
&=& \langle  b^\dagger(t+\tau)b^\dagger(t) b(t)b(t+\tau) \rangle \nonumber \\
&& + \, \langle 
b^\dagger(t+\tau)[b(t+\tau),b^\dagger(t)] b(t) \rangle  \nonumber \\
&=& \langle : I(t+\tau) I(t) :\rangle + \langle I(t) \rangle \delta(\tau),
\label{nopsn}
\end{eqnarray}
where the $\langle:$ $:\rangle$ denotes normal ordering as usual 
\cite{WalMil94}.

The final term in Eq.~(\ref{nopsn}), which is 
proportional to $\delta(\tau)$, is the shot noise. It
 is present because the field is really composed of discrete 
quanta (photons) or particles (atoms). Although 
this is an irreducible quantum fluctuation in the field intensity, I 
will ignore it for now and concentrate on the normally-ordered 
fluctuations. As will be seen below, the shot-noise is (perhaps 
surprisingly) more 
conveniently analysed when considering phase fluctuations. The 
normally-ordered fluctuations are analogous to  
fluctuations in a classical field but their probability distribution 
function, the Glauber-Sudarshan $P$-function, may be non-positive 
\cite{Sud63,Gla63a,Gla63b}. Consideration of these fluctuations leads 
to the third requirement on a laser that
\begin{description}
	\item[(3)]  The output intensity fluctuations are small in the sense 
	that $\forall \tau \; |\langle:I(t+\tau),I(t) :\rangle| \ll \langle I \rangle^{2}$.
\end{description}
Here I am using the notation
\begin{equation}
\langle:I(t+\tau),I(t) :\rangle = \langle: I(t+\tau)I(t):\rangle - 
\langle I \rangle^{2}.
\end{equation}

\subsubsection{Second-order coherence}

The above condition can be recast in terms of Glau\-ber's normalized 
second-order 
coherence function \cite{Gla63a}
\begin{equation}
g^{(2)}(\tau) = {\langle:I(t+\tau)I(t) :\rangle}/{\langle I \rangle^{2}}
\end{equation}
as 
\begin{equation}
|g^{(2)}(\tau) - 1| \ll 1.
\end{equation}
For light which is second-order coherent, $g^{(2)}(\tau) = 1$. Thus, 
the condition {\bf (3)} can be restated as: the output of a laser is 
approximately second-order coherent. This can be contrasted with the 
light from a thermal source, for which $g^{(2)}(0)=2$ \cite{WalMil94}.

The interpretation of $g^{(2)}(\tau)$ is that it gives the relative 
change in the likelihood that a boson (photon or atom) will be 
observed a time $\tau$ later than a boson-detection. Explicitly, the 
probability for observing a boson in the interval 
$(t+\tau,t+\tau+dt)$ given that one was observed at time $t$ is 
$g^{(2)}(\tau)\langle I \rangle dt$. In an approximately 
second-order coherent beam, the boson arrival times are 
approximately independent. In an exactly second-order coherent beam 
the arrival times would be Poissonian. Thermal light, with 
$g^{(2)}(0)=2$ is super-Poissonian. This bunching has 
recently been confirmed for a thermal source of Bose atoms as well 
\cite{YasShi96}.

It might be thought that 
the definition adopted here would also exclude sub-Poissonian light 
with $g^{(2)}(0)-1<0$, as 
produced by a regularly-pumped laser for example 
\cite{MarZol89,HaaTanWal89,KhaKog90,RalSav91,RitZolGarWal91}. 
However, this is not 
the case because the arrival of a photon in this case has only a very 
small negative influence on the arrival likelihood multiplier 
$g^{(2)}(\tau)$. Specifically, if $\bar{n} \gg 1$ is the mean number of photons 
in the laser cavity then $g^{(2)}(\tau)-1 > - 1/(2\bar{n})$.

\subsubsection{The noise spectrum}

Another common way that intensity noise is quantified is by the 
intensity noise spectrum
\begin{equation} 
S(\omega) = {\langle I \rangle^{-1}} \int\!d\tau e^{i\omega 
\tau}\langle I(t+\tau),I(t) \rangle.
\end{equation}
The normalization here is chosen so that the shot noise contribution 
is equal to one:
\begin{equation} \label{ins}
S(\omega) = 1 + \langle I \rangle \int\!d\tau e^{i\omega 
\tau}[g^{(2)}(\tau)-1].
\end{equation}
For an exactly second-order coherent beam the shot noise is the only 
contribution. As will be seen, the intensity noise spectrum shows an 
enormous difference between a laser and a thermal source with the 
same intensity and spectral width.

\subsection{Well-defined phase}

If the laser output has small intensity fluctuations then the only 
significant variation in its complex amplitude is that due to phase 
fluctuations. This implies that the first order coherence function
\begin{equation} \label{g1}
G^{(1)}(\tau) = \langle b^\dagger(t+\tau)b(t) \rangle,
\end{equation}
or its normalized form
\begin{equation}
g^{(1)}(\tau) = \langle b^\dagger(t+\tau)b(t) \rangle/\langle b^\dagger b 
\rangle,
\end{equation}
is a useful measure of phase fluctuations. It also has the advantage 
that it is sensitive to phase fluctuations without requiring the 
existence of a mean field having a well-defined phase. Unlike the 
field itself $b(t)$, the bilinear 
combination in Eq.~(\ref{g1}) is measurable even for an atom field.
For $\tau=0$, it is simply the mean intensity $\langle I \rangle$, but 
as $\tau$ increases $|G^{(1)}(\tau)|$ decreases as the phase of the 
field gradually becomes decorrelated from its value at time $t$. 
As $\tau \to \infty$, $|G^{(1)}(\tau)| \to 0$. By contrast, an approximately
 first-order coherent source would satisfy $|g^{(1)}(\tau)|
\approx 1 \, \forall \tau$. This clearly cannot be imposed as a 
condition on any finite laser.

If we cannot expect $|G^{(1)}(\tau)|$ to remain constant for all time, 
we can enquire how long it takes to decay. The characteristic time for 
this decay is known as the {\em coherence time}, and can be simply 
defined as \cite{Gar85}
\begin{equation} \label{deftcoh}
\tau_{\rm coh}= \int_{0}^{\infty} |g^{(1)}(\tau)| dt.
\end{equation}
The coherence time is the time scale over which the phase of the 
field remains roughly constant. If we wish to say that the 
phase of the field is well-defined, then we should be able to measure 
its value and verify that it does remain constant over this time scale. 
Of course we cannot measure the absolute phase of the atom field, and 
we very rarely measure the absolute phase of an optical-frequency 
electromagnetic field. However we can measure the phase relative to 
another source (a laser), which we know to have a very long coherence 
time, by doing an interference experiment. Alternatively, we can 
measure the two-time correlation function $g^{(1)}(\tau)$ by 
putting a time-delay in one arm of an interferometer.

If the field were classical then there would be no difficulty with 
making an arbitrarily precise phase measurement in an arbitrarily 
short time. But we are dealing with quantum fields, and the accuracy 
of a phase measurement (via interferometry or otherwise) is limited 
by the graininess of the field. This is the problem of shot noise 
which was mentioned above. In order to obtain a
 precise result from a phase measurement, we need to measure a 
 portion of the output field containing a large mean number $\bar{n}$ of bosons. 
 That is because for a phase measurement on any state 
 with a mean number $\bar{n} \gg 1$ of bosons, the minimum 
 uncertainty  scales as \cite{SumPeg90}
\begin{equation}
\delta \phi \agt \bar{n}^{-1},
\end{equation}
while for states with a Poisson distribution of mean $\bar{n}\gg 1$ 
(which is more relevant to a laser)
\begin{equation}
\delta\phi \agt \bar{n}^{-1/2}.
\end{equation}
In either case we clearly require $\bar{n} \gg 1$ to obtain a good 
measurement of phase. Now if the phase measurement lasts a time $T$, 
then the mean number of 
bosons available is $\bar{n}=\langle I \rangle T$. But we also require $T$ to be 
much less than the coherence time $\tau_{\rm coh}$, otherwise the 
phase will diffuse over the course of the measurement. Thus we have the 
following condition on the laser output:
\begin{equation} \label{lr}
\langle I \rangle \tau_{\rm coh} \gg 1.
\end{equation}

This can be stated very concisely using $G^{(1)}(\tau)$ as
\begin{description}
	\item[(4)]  The output phase fluctuations are small in the sense 
	that $\int \!d\tau\,|G^{(1)}(\tau)| \gg 1$.
\end{description}
To my knowledge a condition such as this has never been proposed 
before, so in that sense this final 
condition is the most important of the four presented here. This 
condition is easily satisfied by all optical lasers, as will be 
explained below. It is only with 
lasers which have been deliberately filtered that 
the shot noise becomes a serious limitation to whether an interference 
pattern can be observed \cite{HarSan96}. This filtering has been done to 
prove that spatial interference exists even when there is at most one 
photon in the apparatus at any given time \cite{PflMan68}, or temporal 
interference when there is at most one photon arrival per beat period 
\cite{HarBroFujSan95}.

\subsubsection{Power Spectrum}

The coherence time introduced here is basically the reciprocal of the 
spectral width $\delta\omega$. The latter is usually defined as the 
full-width at half-maximum height of the 
power spectrum
\begin{equation}
P(\omega) = \frac{1}{2\pi}\int_{-\infty}^{\infty} \!d\tau\, e^{i\omega\tau} 
G^{(1)}(\tau),
\end{equation}
which is defined so that $\int_{-\infty}^{\infty} d\omega 
P(\omega)  = \langle I \rangle$.
For many sources
the power spectrum is Lorentzian:
\begin{equation} \label{lor}
P(\omega) = (2\pi)^{-1} \frac{\langle I \rangle \delta\omega}
{(\omega-\bar{\omega})^2+(\delta\omega/2)^2} .
\end{equation}
This gives an alternative statement of the fourth condition, namely 
that the resonant spectral intensity $P(\bar\omega)$ be much greater 
than unity. 

From these considerations we see that the spectral width of a laser must be small in two 
senses. From condition {\bf (2)} we require $\delta\omega \ll \bar\omega$ and 
from condition {\bf (4)} we require $\delta\omega \ll \langle I \rangle$. 
For single-mode optical lasers with an output power of greater than about 
one milliwatt, $\langle I \rangle > \bar\omega$, so 
condition {\bf (4)} is actually weaker than condition {\bf (2)}, and 
is always satisfied. Atom lasers are a different matter; as will be 
discussed in Sec.~V they may have difficulty satisfying condition 
{\bf (4)}. Furthermore, the quantity on the left side of Eq.~(\ref{lr})
cannot be changed by longitudinal acceleration of the laser beam. Although such 
acceleration can increase the monochromaticity of the beam, it 
does so by increasing 
$\bar\omega$ but leaving $\delta\omega$ unchanged, as we saw in Sec.~III~B~2. 
Since it cannot change the flux $\langle I \rangle$ it cannot change the 
ratio $\delta\omega:\langle I  \rangle$.

\subsubsection{Bose statistics}

Although I have derived condition {\bf (4)} from considerations of phase 
measurements, it could also have been derived
 from considerations of another fundamental 
property of a laser output: its constituents must be bosons. None of 
the first three conditions require this but the fourth does as can be 
seen as follows. Consider a section of the laser output of length 
$Z$ much longer than the coherence length $l_{\rm coh}\sim c\tau_{\rm coh}$, 
where $c$ is the mean speed of the bosons. 
Being of finite length, this section can be described by discrete 
longitudinal modes in $k$ space with separation $2\pi/Z$. If the uncertainty in 
the spatial frequency is of order $\delta k \sim \delta\omega/c$, then 
the number of these modes which are significantly populated is of 
order $Z \delta k \sim Z\delta\omega/c$. Now the temporal duration of 
this section is $T=Z/c$, so the mean number of bosons present 
in the section is $\langle I \rangle Z/c$. Thus the mean number of 
bosons per mode is of order $\langle I \rangle / \delta \omega$, which 
is independent of $Z$ and must be much greater than unity according 
to condition {\bf (4)}. That is, condition {\bf (4)} is equivalent to 
the condition that the output field be highly Bose degenerate.

This requirement can 
be understood from the fundamental principle {\bf($\pi$)}. For a quantum field 
to approach a classical field it must be highly excited, with many 
quanta per mode. This is necessary for the quantum fluctuations 
(caused by the discreteness of the quanta) to be comparatively
small. A fermion field (such as describes all of the known fundamental 
constituents of matter) cannot have a classical limit because there can 
be at most one particle per mode (the Pauli exclusion principle). Thus 
while condition {\bf (3)} serves to limit the size of the `classical' fluctuations in 
the field, condition {\bf (4)} does likewise for the quantum fluctuations.

Yet another way to interpret condition {\bf (4)} is as follows. The 
time $\langle I \rangle^{-1}$ is the mean time between one boson and 
the next in the output of the laser. Writing condition {\bf (4)} as 
$\langle I \rangle^{-1} \ll \tau_{\rm coh}$, this implies that
 the output field is still coherent from one boson to the 
next. If instead $\langle I \rangle^{-1} \agt \tau_{\rm coh}$,
then the ``phase'' of one boson emitted by the laser 
would be essentially independent of the ``phase'' of the subsequent ones (I use 
quote marks here because one boson individually can have no phase). 
Stated in this way, condition {\bf (4)} seems to come closest to the 
intuitive conviction that that an atom laser output should be ``coherent''. 
Without condition {\bf (4)}, quantum statistics would play no significant 
role, and there would be no reason to treat the beam as a quantum 
field at all.

\section{Adding bosons, one by one}

In this section I will present an idealized model of a device which 
satisfies the conditions {\bf(1)}--{\bf(4)} and hence can be 
classified as a laser. Consider an optical model consisting of a 
cavity supporting a single mode which is damped from one end mirror. 
This automatically produces an output beam with the desired 
directional properties. The laser gain mechanism required to maintain 
a stationary operation is a scheme which simply adds photons to the 
laser mode, one at a time, at Poisson-distributed times. Part of the 
motivation for presenting this model is to dispell any belief that 
the laser gain mechanism has to be ``coherent'' in some special way.

\subsection{The laser model}

The laser gain is achieved by the passage of
two-level atoms through a cavity. The atoms are resonant with the 
 field, and are initially excited. Each atom is
  put through the cavity repeatedly until it is known to 
 have given up its quantum of energy to the field. The 
interaction Hamiltonian is
\begin{equation} \label{linHam}
H = i \Omega( \sigma a^\dagger -  \sigma ^\dagger a),
\end{equation}
where $a$ is the annihilation operator for the cavity mode, $\sigma = 
|g\rangle \langle e|$ is the lowering operator for the atom, and 
$\Omega$ is the one-photon Rabi frequency. Let the interaction 
time $\tau$ be such that $\epsilon \sqrt{\bar{n}} \ll 1$, where 
$\epsilon=\Omega\tau$ and $\bar{n}$ is the 
mean intracavity photon number. Then the unitary operator $\exp(-iH\tau)$ 
acting on the initially factorized state $\rho \otimes 
|e\rangle\langle e|$ 
can be expanded to second order in $\epsilon$ to give the entangled state 
\begin{eqnarray}
R &=& 
\rho \otimes |e\rangle\langle e|  + \epsilon \left( a^\dagger \rho \otimes 
|g\rangle\langle e| + {\rm H.c.} \right) \nonumber \\
&& + \, \epsilon^2 \left(
a^\dagger \rho a \otimes |g\rangle \langle g| - \smallfrac{1}{2}  
\{ a a^\dagger, \rho \} \otimes |e\rangle \langle e| \right).
\end{eqnarray}

If the outgoing atom is detected in the excited state, then the conditioned 
state of the field (the norm of which represents the probability of this 
detection result) is, to first order in $\epsilon^{2}$,
\begin{eqnarray}
\tilde{\rho}_e &=& \bra{e}R\ket{e} \,
 =\, \left(1 - \epsilon^2{\cal A}[a^\dagger] \right) \rho  \nonumber \\
 &=& \exp(-\epsilon^2 
a a^\dagger /2)  \rho \exp(-\epsilon^2 a a^\dagger/2),
\end{eqnarray}
where the  superoperator ${\cal A}[c]$ is defined for an arbitrary 
operator $c$ by
\begin{equation}
{\cal A}[c]\rho = \smallfrac{1}{2}\{ c^\dagger c, \rho\}.
\end{equation}
If the atom is detected in the ground state (which happens rarely), the 
state is
\begin{equation}
\tilde{\rho}_g = \bra{g}R\ket{g} = \epsilon^2 {\cal J}[a^\dagger] \rho,
\end{equation}
where the superoperator ${\cal J}[c]$ is defined by
\begin{equation}
{\cal J}[c]\rho = c\rho c^\dagger.
\end{equation}

If the atom is detected in the ground state, then the field has 
gained a photon and the process can stop. If it is detected in the 
excited 
state, one must try again with the same 
atom (or, more realistically, another excited atom). Say $K$ 
atoms are required before the $(K+1)$th is detected in the ground 
state. The unnormalized state after the $(K+1)$th atom is 
\begin{equation}
\tilde{\rho}_K = \epsilon^2 {\cal J}[a^\dagger] \exp[-K\epsilon^2 a 
a^\dagger /2]  \rho \exp[-K\epsilon^2 a a^\dagger/2].
\end{equation}
The norm of this density operator is equal to the probability that this 
many atoms are needed. Thus, the average density operator, given that 
an atom is finally detected in the ground state, is
\begin{equation} \rho' = \sum_{K=0}^{\infty} \tilde{\rho}_K.
\end{equation}
Using the fact that $\epsilon^{2}$ is small, the sum can be converted to an 
integral by setting $u=\epsilon^2 K$:
\begin{equation}
\rho' ={\cal J}[a^\dagger] \int_0^{\infty} \exp(-u a a^\dagger /2)
 \rho \exp(-u a a^\dagger/2) du.
\end{equation}
This can be formally evaluated as 
\begin{equation}
\rho' = {\cal J}[a^\dagger] {\cal A}[a^\dagger]^{-1} \rho.
\end{equation}
The superoperator ${\cal A}[a^\dagger]^{-1}$ is well-defined 
because $a a^\dagger$ is a strictly positive operator \cite{seeBraCav94}.

The action of the superoperator ${\cal J}[a^\dagger]{\cal A}[a^\dagger]^{-1}$ is 
to add a photon to the system irrespective of its initial state. That 
is to say, it shifts the photon number distribution upwards by one. If 
this addition of a photon is assumed to occur at Poisson-distributed 
times, then we can obtain a Markovian master equation for the field. 
If we also include the 
linear damping at rate $\kappa$, and let the gain (the rate of photon 
addition) be $\kappa\mu$, then we get
\begin{eqnarray} \label{lme}
\kappa^{-1}\dot{\rho} &=& \mu\left({\cal J}[a^\dagger]{\cal 
A}[a^\dagger]^{-1}-1\right)\rho + {\cal D}[a]\rho \nonumber \\
&=& \mu{\cal D}[a^\dagger]{\cal A}[a^\dagger]^{-1}\rho + {\cal D}[a]\rho. \label{551a}
\end{eqnarray}
 Here I am using another superoperator ${\cal 
D}[c]$ defined for an arbitrary operator $c$ by
\begin{equation}
{\cal D}[c]\equiv {\cal J}[c] - {\cal A}[c].
\end{equation}
It can be verified that the ideal laser 
master equation (\ref{lme}), first derived in 
Ref.~\cite{Wis93a}, is of the 
required Lindblad form \cite{Lin76}.

\subsection{The laser statistics}

This master equation is conveniently solved using the Glauber-Sudarshan
$P(\alpha,\alpha^{*})$ representation 
\cite{Sud63,Gla63a,Gla63b,Gar91,WalMil94}. Defining number and 
phase variables in terms of the complex amplitudes
\begin{equation}
	n=|\alpha|^2 \; ; \;\; \varphi = {\rm Im} \log \alpha,
\end{equation}
and using the operator correspondences for the $P$ function, 
one can demonstrate the following superoperator correspondences:
\begin{eqnarray}
	{\cal A}[a^\dagger]\rho & \to & n\left(1-\partial_n\right) P(n,\varphi),
	\nonumber 	\\
	{\cal D}[a^\dagger]\rho & \to & \left[ -\partial_n  n 
	\left(1-\partial_n\right) + \frac{1}{4n}\partial_\varphi^2 
	\right] P(n,\varphi) ,\nonumber \\
	{\cal D}[a]\rho &\to & \partial_{n} n P(n,\varphi).
\end{eqnarray}
Putting these into Eq.~(\ref{551a}), and dropping terms of order $1/n^2$ 
and smaller gives the Fokker-Planck equation
\begin{equation}
	\kappa^{-1}\dot{P} = \left( - \partial_n[\mu - n] + \frac{\mu}{4n^2}
	\partial_\varphi^2 \right) P.
	\label{551g}
\end{equation}

The first notable feature of this equation is that there is no 
diffusion in the intensity $n$. Thus $n$ relaxes exponentially at rate 
$\kappa$ to its stationary value $\mu$. From the definition of the $P$ 
function, this means that the stationary state of this laser is a 
mixture of coherent states of mean photon number $\mu$. Thus the 
intracavity photon number statistics are Poissonian with mean $\mu$. 
Furthermore, the second-order coherence function is defined in terms 
of the $P$ function variables by
\begin{equation}
g^{(2)}(\tau) = \langle n(t+\tau)n(t) \rangle / \langle n \rangle^{2},
\end{equation}
so for this ideal model we have the stationary result 
$g^{(2)}(\tau)=1 \; \forall \tau$. That is to say, the 
output is exactly second-order coherent, satisfying condition {\bf (3)}.
The mean intensity of the output is given by $\langle I \rangle = 
\kappa \langle n\rangle = \kappa\mu$, which is equal to the gain (as 
it must be at steady-state).

The stationary value $n=\mu$ also allows us to replace $n$ in the 
diffusion term of (\ref{551g}) by $\mu$. 
Then the dynamics of the laser is completely 
determined by the phase diffusion
\begin{equation}
	\dot{P}(\varphi) = \frac{\Gamma}{2} \partial_\varphi^2 P(\varphi),
	\label{551i}
\end{equation}
where $\Gamma$ is the fundamental rate of phase diffusion of the laser, 
given by
\begin{equation}
	\Gamma = \frac{\kappa}{2\mu}.
	\label{551j}
\end{equation}
This expression is well-known and an heuristic derivation is given by Loudon 
\cite{Lou83}. 

Equation (\ref{551i}) implies that $\varphi(t)$ is a Wiener process, 
so that $\langle [\varphi(t+\tau)-\varphi(t)]^{2} \rangle = 
\Gamma\tau$ \cite{Gar85}. We can use this fact to prove that the 
coherence time (\ref{deftcoh}) is equal to $2\Gamma^{-1}$, since in 
steady state
\begin{eqnarray}
|g^{(1)}(\tau)| &=& \left|\left\langle 
\sqrt{n(t+\tau})e^{-i\varphi(t+\tau)}\sqrt{n(t)}e^{i\varphi(t)} 
\right\rangle\right|/ \langle n \rangle \nonumber \\
&=& \left\langle e^{-i[\varphi(t+\tau)-\varphi(t)]}\right\rangle \nonumber 
\\ &=& \exp(-\Gamma \tau/2),
\end{eqnarray}
Thus the power spectrum of the laser is Lorentzian as in 
Eq.~(\ref{lor}), with a width $\delta \omega = \Gamma$. Since we must 
have $\kappa \ll \bar\omega$ in order to use the Markovian description of 
damping, it is clear that the condition {\bf (2)} is satisfied with 
$\delta\omega \ll \bar\omega$. Also we have
\begin{equation}
\tau_{\rm coh}\langle I \rangle \sim \langle I \rangle / \Gamma = 2 
\mu^{2},
\end{equation}
so providing that the mean intracavity photon number 
$\mu$ is much greater than one [as was assumed above in obtaining the 
Fokker-Planck equation (\ref{551g})], condition {\bf (4)} is easily 
satisfied.

\subsection{Comparison with linear amplifier}

The notable feature of the laser gain mechanism presented above is 
that is adds photons one by one, independent of the state. This is to 
be contrasted with a linear amplifier, in which the rate of addition 
of photons is proportional to the number of photons plus one, as 
usual for stimulated emission. The 
master equation for a linear amplifier with the same damping term and 
the same mean photon number as the above laser is
\begin{equation} \label{linamp}
\kappa^{-1}\dot\rho = \mu{\cal D}[a^\dagger](\mu + 1)^{-1}\rho + {\cal 
D}[a]\rho
\end{equation}
This is best converted into a Fokker-Planck equation for the $P$ 
function in terms of $\alpha$ and $\alpha^{*}$ \cite{WalMil94}
\begin{equation}
\dot{P} = \frac{\kappa}{\mu+1}\left[ 
\smallfrac{1}{2}\partial_{\alpha}\alpha + 
\smallfrac{1}{2}\partial_{\alpha^{*}}\alpha^{*} + 
\mu\partial_{\alpha}\partial_{\alpha^{*}}\right]P.
\end{equation}
From this it is easy to show that the stationary first order 
correlation function is
\begin{eqnarray}
|g^{(1)}(\tau)| &=& \left|\left\langle 
\alpha^{*}(t+\tau)\alpha(t)\right\rangle\right|/ \langle |\alpha|^{2} \rangle \nonumber \\
&=& \exp\left(-\frac{\kappa\tau/2}{\mu+1}\right).
\end{eqnarray}
For $\mu\gg 1$, the spectral width is $\delta\omega = \kappa/(\mu+1) \approx 
\kappa/\mu$, which is only twice that of a laser (\ref{551j}) with the same 
intensity $\langle I \rangle = \kappa \mu$. Clearly conditions {\bf 
(2)} and {\bf (4)} are still satisfied for $\mu \gg 1$.

The factor of two difference in the linewidth is
because this source has amplitude fluctuations as well as phase 
fluctuations contributing to the decorrelation of the field. It is 
these amplitude fluctuations which  
exclude the linear amplifier from the class of lasers. We find in 
this case
\begin{equation} \label{g2la}
g^{(2)}(\tau) = 1+|g^{(1)}(\tau)|^{2}=1 + \exp[-\kappa\tau/(\mu+1)],
\end{equation}
which clearly clearly violates condition {\bf (3)}.
The difference in the intensity fluctuations is even more striking in 
the low-frequency intensity noise spectrum (\ref{ins}). For the laser 
model above we have
\begin{equation}
S_{\rm laser}(0) = 1,
\end{equation}
representing shot noise only. For the linear amplifier,
\begin{equation}
S_{\rm lin.amp.}(0) = 1+2\mu(\mu+1),
\end{equation}
which is enormously far above the shot-noise level for $\mu \gg 1$. 

The result obtained here for a linear amplifier is identical to what 
would be obtained by passively filtering a beam emitted by a 
source of bosons in thermal equilibrium. The source would have to be
extremely Bose degenerate, with $\mu$ bosons per mode, and the filter 
would have to block all except a narrow range of 
energies $\hbar\delta\omega = \hbar\kappa/(\mu+1)$. As well as failing 
to satisfy condition {\bf (3)}, this method of producing an intense
monochromatic beam is very inefficient, in that it filters out almost 
all of the input. As pointed out by Holland {\em et al.} \cite{Hol96}, 
this is in contrast to 
a laser in which Bose statistics enhance the transfer of bosons 
into the laser mode.

\subsection{Is stimulated emission necessary?}

Since the ``stimulated emission of radiation'' is part of the acronym 
for laser, it might be thought that stimulated emission is essential 
to produce a laser. While this is true for a typical 
laser (as will be examined in Sec.~V), the fact that the model of 
Sec.~IV~A adds photons one by one suggests that it is not true in general. 
I will now show that stimulated emission is indeed {\em not} necessary 
for laser action. 

Stimulated emission is a simple 
consequence of the linear coupling of the laser field to its source, 
as in Eq.~(\ref{linHam}). That is to say, the Hamiltonian 
(\ref{linHam}) is linear in the annihilation operator $a$ which, for 
classical fields, can be replaced by the $c$-number $\sqrt{n}e^{i\phi}$. 
Whenever a rate is calculated in quantum theory 
it depends on the square of the Hamiltonian. Hence the fundamental 
gain rate from a linear coupling will vary as $n$, which is the 
so-called stimulated emission or Bose-enhancement factor. A fully quantum 
calculation of course gives spontaneous emission as well, and hence a 
gain rate proportional to $n+1$.

Since stimulated emission can be traced to the presence of $a$ in the 
coupling Hamiltonian, the only way to remove it is to substitute for 
$a$ a different 
lowering operator, one which does not increase with $n$. That 
is to say, in Eq.~(\ref{linHam}), we replace
\begin{equation}
	a = \sum_{n=1}^{\infty}\sqrt{n}\ket{n-1}\bra{n} 
\end{equation}
by the Susskind-Glogower \cite{SusGlo64} $e\equiv \widehat{e^{i\phi}}$ operator
\begin{equation}
e = 
	a (a^{\dagger} a)^{-1/2} = \sum_{n=1}^{\infty} \ket{n-1}\bra{n}.
\end{equation}
The new Hamiltonian would be extremely 
nonlinear if expressed in terms of $a$ and $a^{\dagger}$, and would 
be quite impossible to realize physically, but it cannot be 
denied that it will not exhibit any stimulated emission. The above 
derivation will follow through, and indeed will be simplified by the 
fact that $e e^{\dagger} = 1$. Thus in place of 
Eq.~(\ref{lme}) we will have 
\begin{equation}
    \kappa^{-1}\dot\rho = 
    \mu{\cal D}[e^{\dagger}]\rho + {\cal D}[a]\rho
	\label{lmetilde}.
\end{equation}
It can be shown that this master equation produces exactly the same 
photon number statistics as does Eq.~(\ref{lme}). Moreover, for $\mu 
\gg 1$, the linewidth 
of the laser described by Eq.~(\ref{lmetilde}) will be half 
that of the laser described by Eq.~(\ref{lme}). Thus, all of the 
conditions for the device to be considered a laser can be satisfied 
without there being any stimulated emission at all. However, as 
stressed above, this is a completely artificial model and in real 
lasers (as will be analysed in the next section), the Bose 
enhancement factor is essential for their operation above threshold.

\section{A Generic Laser Model}

The laser model presented in the preceding section proves 
that it is possible to construct a device satisfying 
the criteria {\bf (1)}--{\bf (4)} simply by adding bosons one at a 
time to a damped mode. However, it has two shortcomings: first, its 
properties are too idealized; second, its dynamics are not typical of 
a laser. For these reasons I will present in this section a more 
realistic model of a laser which will illustrate the typical features 
of laser dynamics. I suggest that the following list contains the most 
important of such features:
\begin{enumerate}
\item a {\em cavity} supporting the {\em laser mode} of the boson 
field and allowing an {\em output beam} to form.
\item a {\em source} of bosons which is {\em coupled} to the laser mode.
\item an {\em irreversibility} which favours the transfer of bosons 
from source to laser mode. 
\item a {\em sink} which also takes bosons from the source. 
\item a {\em pump} to replenish the source.
\end{enumerate}
As we will see, a crucial requirement for the device to operate as a 
laser is that the transfer of bosons from source to laser mode be at 
least comparable to the transfer from source to sink.

\subsection{The dynamical model}

Before discussing the realization of the above laser 
dynamics in specific physical systems, I wish to present a generic 
model which can be analysed in these terms. For simplicity I will 
model the laser mode and the source as two different modes of the 
boson field with annihilation operators $a$ and $c$ respectively. In 
addition there is another mode $b$ (not necessarily of the same field) 
which will cause the 
irreversibility in the coupling between $a$ and $c$. The dynamics of 
the total system is then governed by the following master equation
\begin{equation} \label{startme}
\dot{W} = \sum_{i=1}^{5}{\cal L}_{i}W,
\end{equation}
where the five Liouville superoperators corresponding to the five 
features listed above are
\begin{eqnarray}
{\cal L}_{1}W &=& \kappa{\cal D}[a]W \\
{\cal L}_{2}W &=& -i[g( c^\dagger a b + c a^\dagger b^\dagger ),W ] \\
{\cal L}_{3}W &=& \lambda {\cal D}[b]W \\
{\cal L}_{4}W &=& \gamma (N+1){\cal D}[c] W\\
{\cal L}_{5}W &=& \gamma N {\cal D}[c^\dagger] W
\end{eqnarray}

The first superoperator ${\cal L}_{1}$ describes damping of the laser 
mode at rate $\kappa$, as in the model of the preceding section. The 
second ${\cal L}_{2}$ is a Hamiltonian (reversible) coupling of 
 the source mode $c$ to the laser mode 
$a$ and another mode $b$. The third ${\cal L}_{3}$ damps this other 
mode $b$, and $\lambda$ will be assumed very large so that the 
coupling ${\cal L}_{2}$ is effectively made unidirectional. The 
fourth and fifth together ${\cal L}_{4}+{\cal L}_{5}$ describe the 
coupling of the source to a broad-band reservoir with $N$ bosons per 
mode. The pump term (${\cal L}_{5}$) is the usual linear 
amplifier while the sink term (${\cal L}_{4}$) is the usual linear 
loss.

We wish to simplify this three-mode master equation into a 
single-mode master equation by eliminating first mode $b$ and then 
mode $c$. Under the assumption that $\lambda \gg g$ we can follow the 
adiabatic elimination 
method of Refs.~\cite{WisMil93a} and \cite{Wis93a} to derive a master 
equation for $R$, the state matrix for modes $a$ and $c$ alone:
\begin{equation} \label{meR}
\dot{R} = [{\cal L}_{1}+{\cal L}_{e} + {\cal L}_{4} + {\cal L}_{5}]R,
\end{equation}
where the new superoperator replacing ${\cal L}_{2}$ and ${\cal 
L}_3$ is 
\begin{equation}
{\cal L}_{e}R =\eta{\cal D}[c a^\dagger]R,
\end{equation}
where $\eta = g^2/\lambda$. 
This is the (now irreversible) coupling between the source and laser 
mode. In order to adiabatically eliminate mode $c$ as well we assume 
that $N\ll 1$. This is a natural assumption to make as Bose-degenerate beams 
($N \agt 1$) are difficult to come by unless one already has a 
laser. Under this assumption we can expand the state matrix $R$ as
\begin{equation} \label{expR}
R = \rho_{0}\otimes\ket{0}\bra{0} + \rho_{1}\otimes\ket{1}\bra{1},
\end{equation}
where the $\rho_0$ and $\rho_1$ are density operators for the laser 
mode $a$, and the number 
states are those for the source mode $c$. We will see that in 
steady-state $\rho_{1}\sim N \ll 1$ so that $\rho \simeq \rho_{0}$.

Substituting the expansion (\ref{expR}) into the master equation 
(\ref{meR}) and using $N \ll 1$ yields
\begin{eqnarray}
\dot{\rho}_{0} &=& \left({\cal L}_{1} - \gamma N \right) \rho_{0} 
+ \left(\gamma + \eta{\cal 
J}[a^\dagger]\right)\rho_{1} \label{rho0} \\
\dot{\rho}_{1} &=& \left({\cal L}_{1}-\eta{\cal 
A}[a^\dagger]-\gamma\right)\rho_{1}+\gamma N \rho_{0} \label{rho1}
\end{eqnarray}
Assuming that $\gamma \gg \kappa$, we can neglect the ${\cal L}_{1}$ 
term in the equation for $\rho_{1}$ and, since $\rho_{0}\simeq \rho$ 
varies on the time scale $\kappa$ (as will be shown), we can assume 
that $\rho_{0}$ is constant on the time scale that $\rho_{1}$ relaxes.
Thus we can slave $\rho_{1}$ 
to $\rho_{0}$ by calculating the stationary state of Eq.~(\ref{rho1}):
\begin{equation}
\rho_{1} \simeq \gamma N\left(\gamma+\eta{\cal A}[a^\dagger]\right)^{-1} \rho_{0},
\end{equation}
which is of order $N$ as promised. Substituting this back into 
Eq.~(\ref{rho0}) yields
\begin{equation}
\dot\rho_{0} = \left[{\cal L}_{1} + \gamma N \left(\gamma + \eta{\cal 
J}[a^\dagger]\right)\left(\gamma + \eta{\cal A}[a^\dagger]\right)^{-1} - 
\gamma N \right]\rho_{0}.
\end{equation}
Since $\rho \simeq \rho_{0}$ we can rewrite this as the following master 
equation for the laser mode alone:
\begin{equation} \label{lasme2}
\kappa^{-1}\dot\rho =  {\cal D}[a]\rho + \nu {\cal D}[a^\dagger]\left( n_{\rm 
s} + {\cal A}[a^\dagger]\right)^{-1} \rho,
\end{equation}
where $\nu=\gamma N / \kappa$ is a dimensionless gain parameter and $n_{\rm 
s}=\gamma/\eta$ is a saturation boson number (also dimensionless). 
Both are typically much greater than unity, and this will be 
assumed in all that follows. As promised, the 
characteristic rate of evolution of $\rho$ is $\kappa$. It can also 
be verified that this master equation (\ref{lasme2}) is of the 
Lindblad form.

\subsection{The laser threshold}

The most important aspect in which Eq.~(\ref{lasme2}) is a more realistic 
laser master equation than Eq.~(\ref{551a}) is that it exhibits a 
threshold. If we define a threshold parameter 
\begin{equation}
\theta = \nu/n_{\rm s},
\end{equation}
then the threshold is at $\theta=1$. However, for $n_{\rm s}$ finite 
this threshold is not infinitely sharp, as I explain in this section.

\subsubsection{Above threshold}

For the laser to be above 
threshold we require $\theta$ not just greater than one, but finitely 
greater than $1$. By this I mean that $\theta - 1$ must 
be a positive number not much 
less than one. Using the $P(n,\varphi)$ function as in Sec.~IV, 
it can then be shown that the mean intracavity boson number 
is, to a very good approximation, given by $\bar{n}= \nu - n_{\rm s}=
n_{\rm s}(\theta - 1)$, and the variance by $\nu = n_{\rm s}\theta$. 
The second-order coherence function is given by 
\begin{equation} \label{516}
g^{(2)}(\tau) = 1 + \frac{\exp[-\kappa(1-\theta^{-1})\tau]}
{n_{s}(\theta - 1)^{2}}.
\end{equation}
From this it is apparent that the condition $\theta$ finitely $>1$ is 
necessary to satisfy condition {\bf (3)}. In fact we require 
\begin{equation} \label{rts3}
\theta - 1 \gg n_{\rm s}^{-1/2}.
\end{equation} 
As for condition {\bf (4)}, 
it can be shown that the rate of phase diffusion is given by
$\Gamma = {\kappa}/({2 \bar{n}})$
so that
\begin{equation} \label{518}
\langle I \rangle / \Gamma = 2 \bar{n}^{2} = n_{\rm 
s}^{2}(\theta-1)^{2}.
\end{equation}
Thus condition {\bf (4)} that $\langle I \rangle \gg \Gamma$ 
is still satisfied without difficulty providing $\theta$ 
finitely $>1$ and $n_{\rm s}\gg 1$. In fact, it is satisfied 
providing $\theta - 1 \gg n_{\rm s}^{-1}$, which is a weaker condition 
than (\ref{rts3}) required to satisfy {\bf (3)}. 

Very far above threshold,
$\theta \gg 1$ with $\nu = n_{s}\theta$ constant, and 
Eqs.~(\ref{516}) and (\ref{518}) go over to the ideal limits derived 
in Sec.~IV with $\mu = \nu$. This limit can be obtained directly from the master equation 
(\ref{lasme2}), for if $n_{s} \ll \bar{n}$ then 
$n_{\rm s}$ can be ignored compared to the superoperator ${\cal 
A}[a^\dagger]$ which is of order $a a^\dagger$. 
Then Eq.~(\ref{lasme2}) goes over to the idealized 
Eq.~(\ref{551a}) with $\mu = \nu$. This is the limit in which the 
stimulated emission is so strong that the gain
 becomes independent of the number of bosons in the laser mode. This 
 is not a paradox, because the state of the source depends on the 
 number of bosons in the laser mode. When the laser boson number is 
 sufficiently large, the source is so depleted that the $n+1$ 
 Bose-enhancement factor is neutralized and the absolutely constant 
 gain of the model of Sec.~IV is reproduced. 
 
\subsubsection{Below threshold}

The below threshold regime is, unsurprisingly, $\theta$ finitely $<1$. 
In this regime the source is, to a first approximation, undepleted, 
and it can be shown that 
the mean boson number is $\bar{n} \approx 
\theta/(1-\theta)$. Therefore, for $\theta> 1/2$, $\bar{n}$ 
will be greater than unity, implying that stimulated emission will 
dominate spontaneous emission. 
But this does not mean that $\theta > 1/2$ puts the laser above 
threshold. The laser will remain below 
threshold as long as $\bar{n} \ll n_{\rm s}$. This translates as the 
condition 
\begin{equation}
1-\theta \gg n_{\rm s}^{-1}.
\end{equation}
Under this condition 
the superoperator  ${\cal A}[a^\dagger]$ (which scales as 
$\bar{n}+1$) will also be 
negligible compared to $n_{\rm s}$. Then Eq.~(\ref{lasme2}) can be 
approximated by
\begin{equation} \label{linapp}
\kappa^{-1}\dot\rho = {\cal D}[a]\rho + \theta{\cal D}[a^\dagger]\rho,
\end{equation}
from which the result $\bar{n}=\theta/(1-\theta)$ is trivial to derive. 
This master equation is of the same form as the linear amplifier (\ref{linamp}),
with $\kappa$ the linear loss and $\theta\kappa$ the linear gain.
However unlike Eq.~(\ref{linamp})
 we have $\theta$ finitely $<1$ whereas $\mu/(\mu+1)$ 
approaches $1$ for $\mu \gg 1$. 

A laser below threshold  fails condition {\bf (3)}, 
since $g^{(2)}(0)=2$ as in 
Eq.~(\ref{g2la}). It will also fail condition {\bf (4)}, as $\langle 
I \rangle = \kappa\theta(1-\theta)^{-1}$ and the linewidth is 
$\Gamma = \kappa(1-\theta)$, so that
\begin{equation}
\int_{0}^{\infty} d\tau G^{(1)}(\tau) \sim \langle I \rangle / 
\Gamma = 
\frac{\theta}{(1-\theta)^{2}}
\end{equation}
is only of order unity for $\theta$ finitely $<1$. The laser 
above threshold satisfies condition {\bf (4)} not only because the 
intensity $\langle I \rangle$ is much greater, but because the linewidth 
$\Gamma$ is much narrower. 

\subsubsection{The significance of the threshold}

The increase in intensity and linewidth-narrowing of the laser above 
threshold are simple consequences of the large mean photon number. 
However the same cannot be said for the change in the intensity 
fluctuations. It is therefore 
instructive to enquire what has changed in the laser dynamics to 
make such a difference to these fluctuations. The requirement for the laser to 
be above threshold is, as we have seen above, that $\theta-1$ is 
positive and not very small. This can be restated as: the mean boson 
number $\bar{n}$ must be not very small compared to the saturation 
boson number $n_{\rm s}$. Returning to the above derivation, this is 
saying that in Eq.~(\ref{rho1}), the damping of $\rho_{1}$ 
proportional to $\Gamma {\cal A}[a^\dagger]$ should not be negligible 
compared to the damping due to $\gamma$. In more physical terms, it 
means that the rate of the transfer of bosons from the source to the 
laser mode should be comparable to the rate of transfer from the 
source to the sink. In other words, the laser mode itself must significantly 
deplete the source, giving a sort of dynamical negative feedback, in order 
for the laser to go above threshold. 

In practical terms, this 
characterization of the threshold transition is equivalent to the 
usual one, that a laser goes above threshold when the linear 
gain exceeds the linear loss. The exponential rise in boson 
number implied by linear gain exceeding linear loss must eventually 
deplete the source, causing the gain to change from being linear to 
being nonlinear. Very far above threshold, where Eq.~(\ref{551a}) 
holds, the gain is so nonlinear that it is a constant.

\subsection{Photon lasers}

The starting master equation (\ref{startme}) used above could 
represent a photon laser based on parametric down conversion. To my 
knowledge, no such laser has been built, but it could be in 
principle. In this laser, the fundamental coupling would be a 
$\chi^{(2)}$ nonlinearity between the three electromagnetic field 
modes $a,b,c$, with 
$\omega_{c} = \omega_{a}+\omega_{b}$. It would be possible to have 
$\omega_{a}=\omega_{b}$, but only if the down-converted modes $a$ and 
$b$ were non-degenerate through their polarization for example. The 
high frequency mode $c$ would  be pumped by a thermal light source 
(driving by another laser is, under the strict conditions established 
above, prohibited if the device is to qualify as a laser itself). 
The cavity for mode $b$ would be low quality ({\em i.e.} would have a short 
lifetime), which is easy to achieve. What would be difficult to achieve 
would be to make the cavity for mode $a$ sufficiently high quality 
that a very large number of photons could build up inside it while 
still satisfying the inequalities assumed above in deriving the final 
master equation (\ref{lasme2}).

In most photon lasers, matter plays a more important role in the gain 
mechanism than simply supplying a nonlinear coupling between optical 
modes. Probably the simplest systems to model are lasers using a 
dilute gas (see for example Ref.~\cite{Wis93a}). 
In this case the source system is the gas of atoms (or molecules). 
A particular transition of the atoms is 
coupled to the laser mode by a resonant electric-dipole 
interaction of the form 
\begin{equation} \label{de}
H = g\left(a^\dagger \ket{1}\bra{2} + a \ket{2}\bra{1}\right).
\end{equation}
 The atoms are pumped by being excited (perhaps by a 
thermal light source such as a flashlamp) into a state from which 
they can make a spontaneous transition into the upper level $\ket{2}$.
This by itself is insufficient, because the interaction (\ref{de}) is 
reversible. What is needed is the analogue of mode $b$, which makes the interaction 
irreversible and favours the transfer of energy to the photon field. 
This role is played by the continuum of electromagnetic field modes 
which cause the level $\ket{1}$ to spontaneously decay. Providing this 
spontaneous decay rate $\gamma_{1}$ is rapid enough, 
any atom which gives up its photon 
to the laser field is immediately transferred to another state from 
which it cannot reabsorb that photon. This gives rise to what is 
 called in gas lasers a situation of population-inversion, but it is apparent 
from the generic model of Sec.~V~A 
that the term ``population-inversion'' is not appropriate for 
describing laser action in general. 

Assuming that the atoms can be adiabatically eliminated, the rate for 
transferring quanta from the atoms to the field is equal to 
$N_{2} g^{2} a a^\dagger/\gamma_{1}$, where $N_{2}$ is the number of 
excited state atoms, and $a a^\dagger = a^\dagger a + 1$ is the number of 
photons in the laser mode plus one. The condition for the laser to be
above threshold is that the number of photons in the laser mode is so 
large that this rate be at least comparable with the rate for other 
processes de-exciting the upper level $\ket{2}$. These other 
processes include spontaneous emission at rate 
$\gamma_{2}$, and collisional de-excitation.

\subsection{Atom lasers}

Being material particles, atoms cannot be created out of energy (as 
can photons and other gauge bosons). 
This means that the source for atoms must be a mode (or set of modes) 
of the atom field. Typically these modes would exist in a trap for 
atoms and the laser mode itself would be the lowest mode of that 
trap (although this is not necessarily so). The 
mechanism which have been proposed 
 for transferring atoms from the source to the laser mode fall into two 
classes: optical cooling 
\cite{WisCol95,SprPfaJanWil95,OlsCasDal95,MoyHopSav96} and 
 evaporative cooling 
\cite{GuzMooMey96,WisMarWal96,Hol96}. These will be discussed
separately below. In both cases atoms can be added to the source by pumping from 
some reservoir, either external to the trap or in the high-energy modes 
of the trap. The removal of the atoms from the laser mode to form a 
coherent output beam is a considerable technical challenge, and would depend on 
the nature of the trap. As stated in Sec.~II, it would be useful for 
the output to be in the form of a guided atomic wave.

\subsubsection{Optical cooling}

Optical cooling is the reduction in the kinetic energy of atoms by 
the transfer of energy and momentum between the atoms, optical laser 
beams, and spontaneously emitted photons. This is the  
mechanism for transferring atoms from the source modes of the trap to 
the laser mode of the trap in the proposed atom lasers of 
Refs.~\cite{WisCol95,SprPfaJanWil95,OlsCasDal95,MoyHopSav96}. 
It is clear that the 
irreversibility in this transfer comes from the spontaneously emitted 
photons. Thus, as in a photon laser with a gaseous gain medium, the 
role of mode $b$ is played by the continuum of electromagnetic field 
modes coupled to a particular atomic transition. One problem with this 
cooling mechanism is that the emitted photons may be reabsorbed by 
other atoms in the laser mode. However this problem could be minimized 
for a laser mode which is linear in aspect (that is, long and 
narrow). This might be achieved by siting the laser cavity within an 
atom waveguide, which would also produce the output in a desirable form.

The condition for threshold in an atom laser of this sort is that the 
depletion of the source due to optical cooling into the laser mode 
(which will of course be proportional to the number of laser mode atoms
plus one) be at least comparable to the depletion of the source due 
to losses from the trap or laser-induced transitions to other trap 
modes. Provided that photon reabsorption is not too great a problem, 
this threshold condition can be achieved, so that the intensity 
fluctuations in the laser output can satisfy condition {(\bf 3)}. To 
satisfy condition {(\bf 4)} it is necessary to calculate the 
linewidth of the the laser. So far this has been done only for the model of 
Ref.~\cite{WisCol95}. Because of unwanted processes, the linewidth 
found in Ref.~\cite{WisCol95} far exceeded the ideal 
limit of Eq.~(\ref{551j}), but was still small enough to satisfy 
condition {\bf (4)}. However, this calculation ignored many 
complicating factors such as inter-atomic collisions, so it remains 
to be seen whether it really would be possible to build an atom laser 
based on this or any other method of optical cooling.

\subsubsection{Evaporative cooling}

Evaporative cooling at first sight appears a more promising route
to an atom laser because unlike optical cooling 
it has already produced Bose condensation of atoms. The basic 
mechanism for evaporative cooling is removal of `hot' atoms 
(evaporation) accompanied by thermalization via collisions. For a cold
dilute gas the dominant form of collisions are binary ones resulting 
from $s$-wave scattering, so the 
coupling Hamiltonian is of the form
\begin{equation} \label{swaveham}
H = \sum_{i\leq j,k\leq l} g_{ijkl} a^\dagger_{i}a^\dagger_{j}a_{k}a_{l},
\end{equation}
where $g_{ijkl}^{*}=g_{klij}$ and $a_{i}$ is the annihilation 
operator for the atom field in the $i$th trap mode. 
To transfer atoms from the source to 
the laser mode requires collisions between two atoms in the source 
modes ($k,l$), putting one atom into the laser mode (say $j=0$) and 
the second atom into some other mode ($i$) with high energy. This 
process will be irreversible, as desired, if the atom in mode $i$ has 
sufficient energy to escape the trap (so that $i$ is really a 
continuum mode). Thus in this case $a_{i}$ plays the role of $b$ and 
$a_{k}a_{l}$ that of the source $c$. For the regime of weak pumping 
($N \ll 1$) the fact that the coupling of the source to the laser mode 
is quadratic in the former causes fairly trivial differences from the 
linear coupling used above \cite{Hol96,WisMarWal96}. However, in the 
other limit ($N \gg 1$), some differences do arise 
\cite{WisMarWal96}; in particular the 
intensity fluctuations are not necessarily Poissonian far above 
threshold. Nevertheless it seems that condition {(\bf 3)} could be satisfied for 
this sort of atom laser providing a suitable output coupler is found.

What is unclear about this sort of laser is whether it could satisfy 
condition {(\bf 4)}. Using the simple model proposed by Holland {\em 
et al.} \cite{Hol96}, it was shown in Ref.~\cite{WisMarWal96} that 
the linewidth of an atom laser based on evaporative cooling would 
typically be much larger than the output flux, so that condition {\bf 
(4)} would not be satisfied. The dominant linewidth broadening 
is due to the fluctuations in the self-interaction energy of the laser 
mode coming from the Hamiltonian term $g_{0000}a^\dagger_{0}{}^{2} a_{0}^{2}$. 
This is linked to the dynamics of the laser because it is the same 
$s$-wave scattering Hamiltonian (\ref{swaveham}) 
which causes both the undesired self-interaction energy of 
the laser mode and the desired coupling between the source and the laser 
modes. 
Thus it may be impossible to build an atom laser using evaporative 
cooling. There are various ways that this conclusion could be avoided. 
First, it may be that by using dipole-dipole collisions, as suggested in 
Ref.~\cite{GuzMooMey96}, the cross-section for self-interactions of 
the laser mode could be made negligible compared to the cross-section 
for source-laser coupling, and so overcome the linewidth problem. 
This might also be achievable by using sympathetic cooling 
\cite{Mya96}, in which another species of atom (which could have a 
large collisional cross-section) is cooled 
evaporatively and takes the desired species (which could have a small 
cross-section) into the Bose degenerate regime through thermalization.
A third possibility is that a more elaborate model of 
standard evaporative cooling based on $s$-wave scattering 
might lead to a different conclusion. All of these suggestions
deserve further investigation.

\section{Summary}

In this paper I have argued that before attempting to build an atom 
laser (an analogue of the familiar photon laser), it is necessary to 
have a rigorous definition of what constitutes a laser. This 
definition should apply equally to photon lasers and to atom lasers. 
The fundamental principle underlying the definition which I propose is that 
\begin{description}
	\item[($\pi$)]  A laser is a device which produces an output field which 
	is well-approximated by a classical wave of 
	fixed intensity and phase.
\end{description}
Thus, the laser itself is treated as a black box; we are only 
interested in its output. The inputs are arbitrary except that they 
cannot themselves be laser fields of the same substrate (atoms 
or photons for an atom or photon laser respectively). 

This 
fundamental principle can be quantified by requiring that the output 
of a laser satisfy four conditions. The first two conditions are 
elementary conditions that can be understood using single-boson 
concepts:
\begin{description}
\item[(1)]{The output is highly directional, and ideally has a single transverse mode.}
\item[(2)] The longitudinal spatial frequency of the output beam has a 
small spread in the sense that $\delta k \ll \bar{k}$.
\end{description}
The second two conditions require many-boson concepts and are most 
easily stated using quantum field theory. Under the first two 
conditions the output field of a laser may be described by 
a field operator $b(t)$ satisfying $[b(t),b^\dagger(t')]=\delta(t-t')$. The 
intensity operator  
$I(t) = b^\dagger(t) b(t)$ measures the output boson flux in units of 
bosons per second. Using these quantities the final two conditions 
can be stated as 
\begin{description}
\item[(3)]  The output intensity fluctuations are small in the sense 
that $\forall \tau \neq 0 \; |\langle I(t+\tau),I(t) \rangle| \ll \langle I \rangle^{2}$.
	\item[(4)]  The output phase fluctuations are small in the sense 
	that $\int\!d\tau\,|b^\dagger(t+\tau)b(t)| \gg 1$.
\end{description}
The fourth condition has, to my knowledge, not been proposed by any 
previous authors and so in some sense is the most important 
contribution of this paper.

As well as defining a laser, I introduced a generic laser model and 
listed five characteristic features which it possesses:
\begin{enumerate}
\item a {\em cavity} supporting the {\em laser mode} of the boson 
field and allowing an {\em output beam} to form.
\item a {\em source} of bosons which is {\em coupled} to the laser mode.
\item an {\em irreversibility} which favours the transfer of bosons 
from source to laser mode. 
\item a {\em sink} which also takes bosons from the source. 
\item a {\em pump} to replenish the source.
\end{enumerate}
The presence of a sink implies that this laser can operate either above or below 
threshold. Only when operating above threshold will it satisfy 
the final two conditions above. For the laser to be above threshold 
requires that the transfer of bosons from source to laser mode be at 
least comparable to the transfer from source to sink. In this regime 
there is a negative correlation between the number of source excitations and the 
number of bosons in the laser mode. This provides the self-regulation 
of intensity which enables condition {(\bf 3)} to be satisfied. By 
contrast, condition {(\bf 4)} may be satisfied for a laser above 
threshold simply because the intracavity boson number is very large.

Finally, I presented a very brief analysis of some of the atom laser 
models which have been proposed, in terms of the above conditions and 
features. These atom laser schemes fall into two broad classes: those 
based on optical cooling of atoms and those based on evaporative 
cooling. The former is closer in operation to the generic laser model 
introduced here, but would be hard to implement because of the 
technical difficulties involved in cooling atoms to extremely low 
temperatures in the presence of optical laser beams. The latter 
appears a more promising route because it has already been used to 
achieve Bose condensation. However, it appears that atom lasers based on 
evaporative cooling would have difficulty satisfying condition {(\bf 
4)}. The identification of this difficulty highlights the importance 
of having a definition for an atom laser.

\acknowledgements

I would like to acknowledge stimulating discussions with M.J.~Collett, 
D.F.~Walls, A-M.~Martins, S.~Dyrting, and many other colleagues.

\appendix
\section*{Symmetries and Conservation laws}

The difference between matter fields and the electromagnetic field
can be seen from the fundamental coupling Hamiltonian 
for quantum electrodynamics (QED), which can be written as
\begin{equation}
	H_{\rm QED} = \sum_{ijk} u_{ijk} \Psi^{\dagger}_{i} ( a_{j}
	 + a^{\dagger}_{j} )\Psi_{k}.
	\label{QED}
\end{equation}
Here $\Psi_{i}$ represents the modes of an electron field,
$a_{j}$ represents  
the annihilation operator for the modes of the photon field,
and $u_{jik}$ is a set of coupling constants. 
This Hamiltonian 
is invariant under the global gauge transformation
\begin{equation}
	\Psi_{i} \to \Psi_{i} e^{if(t)},
	\label{ft}
\end{equation}
which shows that the absolute phase of a matter field is completely 
unobservable, and thus that it is impossible to create a matter field with a 
well-defined absolute phase. This is implicit in the discussion in 
Sec.~II~B about the fact 
that the frequency of a matter field is undefined up to an arbitrary 
additional constant. 
Only the relative phase (and the relative frequency) of one mode of a matter 
field to another mode is physically meaningful. 
The same is not true of the photon  
field because both $a$ and $a^{\dagger}$ appear linearly in 
Eq.~(\ref{QED}) so it is not invariant under the transformation
\begin{equation}
 a_{j} \to a_{j}e^{ig(t)}.
 \label{gt}
\end{equation}
The invariance (\ref{ft}) leads to a conservation law for the total 
number of electrons \cite{ManSha93}
\begin{equation}
	N_{\rm electrons} = \sum_{i} \Psi^{\dagger}_{i}\Psi_{i},
	\label{numpart}
\end{equation}
while the non-invariance (\ref{gt}) 
explains why there is no such conservation law for photon number.

Although there is no absolute conservation law for photon number, there is 
an approximate conservation law at visible frequencies. 
At frequencies $\omega \sim 10^{16} s^{-1}$, 
the characteristic size $d$ of the material 
oscillators (that is, atoms) with these frequencies 
is typically much smaller than the 
wavelength of the resonant Maxwell field. That is, $kd \ll 1$, where 
$k=c/\omega$. Under this condition the coupling 
between atom and the vacuum electromagnetic field can be treated 
using the dipole approximation. This gives a quality factor for the 
oscillator of \cite{ManSha93}
\begin{equation}
	\frac{\gamma}{\omega} = \frac{4}{3}\alpha (kd)^{2},
	\label{Einstein}
\end{equation}
where $\alpha \approx 1/137$ is the QED fine structure constant. 
Because $kd$ is typically of order $10^{-3}$, 
the ratio $\gamma/\omega$ is very small so 
we can also make the so-called rotating 
wave approximation (RWA). This amounts to ignoring non-energy-conserving 
terms in the Hamiltonian (\ref{QED}).

Let us assume for simplicity that each electron is bound in 
a two-level atom. Then the field mode index $i$ can be replaced by 
one letter, $u$ or $l$, corresponding to the electron being in the upper 
or lower electronic level, plus another index $\alpha$ for the 
motional mode of the atom as a whole. 
Then the RWA is effected by  
replacing the fundamental QED Hamiltonian density (\ref{QED}) 
by the approximate Hamiltonian 
\begin{equation}
	H_{\rm RWA} = \sum_{\alpha j \beta} 	v_{\alpha j \beta} 
	\left[ \Psi_{{u},\alpha}^{\dagger} a_{j} \Psi_{{l},\beta} + 
	\Psi_{{l},\alpha}^{\dagger} a_{j}^{\dagger} \Psi_{{u},\beta} \right],
	\label{QEDapp}
\end{equation}
where $v$ is a set of coupling constants related to the set $u$.
This Hamiltonian {\em is} invariant under the transformation 
(\ref{gt}), 
providing it is combined with
\begin{equation}
	\Psi_{{u},\alpha} \to \Psi_{{u},\alpha} e^{ig(t)}
	\label{gt2}
\end{equation}
for the same function $g(t)$. This leads to a conservation law for 
{\em excitation} number, that is, the total number of photons and 
excited state atoms:
\begin{equation}
N_{\rm excitations} = \sum_{\alpha j} \left[
\Psi_{{u},\alpha}^{\dagger}\Psi_{{u},\alpha} + a^{\dagger}_{j} a_{j} \right].
\end{equation}

 This conservation law can be generalized for more complicated atoms. 
It implies that 
under usual optical conditions it is neither possible to measure 
the absolute phase of the electromagnetic field, nor is it possible 
to create a field with a well-defined phase. One could thus conclude that 
the mean field of an optical laser is as much a myth as the non-zero 
mean value of a matter field, and this is the view taken in Ref.~\cite{Mol96}.
I prefer to stress that there is an in-principle 
difference between matter fields and 
the electromagnetic field. After all, the 
RWA is only an approximation and in particular it does not 
apply for the {\em gedankenexperiment} described in the text, in which 
the Lorentz force on a free electron is used 
to measure the temporal phase of an optical standing wave. Similarly, 
as noted in Ref.~\cite{Mol96}, a free-electron laser is a source which 
may produce visible light with a  
temporal phase which could, in principle, be predicted.


\begin{references}
\bibitem{ign}
Here and elsewhere I am assuming that the output atomic beam is 
dilute. That is, that the interatomic separation is sufficiently 
large that interactions between atoms can be neglected. This 
assumption is made in order to keep the formalism tractable, and at 
least in one proposal \cite{WisCol95} it would definitely be satisfied.
\end{references}
\end{document}